\documentclass[a4paper,12pt]{article}

\usepackage{amsmath,amssymb}
\usepackage[applemac]{inputenc}
\usepackage{graphicx}
\usepackage{axodraw}
\usepackage{amsfonts}
\usepackage{bbm}
\usepackage{bm}
\usepackage{mathrsfs}

\input{pix.sty}

\begin{document}


\begin{titlepage}
\begin{flushright}
MS-TP-09-15 \vspace*{1cm}
\end{flushright}
\begin{centering}
\vfill

{\Large{\bf On quarkonium in an anisotropic quark gluon plasma}} 

\vspace{0.8cm}

Owe Philipsen\footnote{ophil@uni-muenster.de} and
Marcus Tassler\footnote{marcus.tassler@uni-muenster.de}

\vspace{0.8cm}
 
{\em
Institut f\"ur Theoretische Physik, Westf\"alische Wilhelms-Universit\"at M\"unster, \\
D-48149 M\"unster, Germany}

\vspace*{0.8cm}

\mbox{\bf Abstract}
 
\end{centering}

\vspace*{0.3cm}
 
\noindent
We reconsider a plasma with an anisotropy imposed on the momentum distribution of the system and study the real time static potential for quarkonia. The distribution function is normalised so as to preserve the particle number in an ideal gas, as required in the Keldysh-Schwinger formalism. In contrast to recent findings without this normalisation, a weak anisotropy does not lead to an increase in the melting temperature for bound states.  
To test for the maximal effect, we also investigate a gluonic medium in the limit of an asymptotically strong anisotropy. The spectral function of quarkonium is calculated for this case and found to be in remarkable  agreement with the corresponding results for an isotropic medium.

\vfill

 
\vfill

\end{titlepage}


\section{Introduction}

The QCD-plasma with an anisotropic momentum distribution 
has been subject to several recent investigations as a 
simple model for the early stages of heavy ion collisions
\cite{Romatschke:2003ms,Romatschke:2004jh,Mrowczynski:2004kv}.
Such a system is obtained by replacing the isotropic momentum space distribution
functions by
\be
f(\bm{k},\xi)=N(\xi)\,f_{\rm iso}
\left(\sqrt{\bm{k}^2
+\xi (\bm{k}\cdot \bm{n} )^2
} \right),
\label{dist}
\ee
which removes modes with momentum components along the direction of the 
anisotropy, $\bm{n}$. Here, $\xi$ parametrises the strength of the anisotropy and
$N(\xi)$ is a normalisation factor with $N(\xi=0)=1$.
Interest has particularly focused on the question how such an
anisotropy influences the dissociation of non-relativistic
quarkonium systems within potential model studies 
\cite{Dumitru:2007hy,Dumitru:2009ni,Burnier:2009yu}.  
The general approach in those works was to consider small anisotropies $\xi\leq 1$
and to calculate the correction to the propagators and the static potential to first order in $\xi$.
The effect of the anisotropy on the normalisation was neglected, i.e.~$N(\xi)\equiv 1$ in those works, and with the exception of \cite{Dumitru:2009fy} all employ the
equilibrium Kubo-Martin-Schwinger (KMS) condition, which strictly speaking 
does not hold in the anisotropic case.
The general conclusion of those investigations was 
that the anisotropy tends to decrease the effect of Landau damping and 
thus to increase the dissociation temperature.

In this paper we assess the extent to which these results are modified by 
taking a $\xi$-dependent normalisation factor into account and solving
again the Dyson-Schwinger equations for the non-equilibrium system. 
This is motivated by the fact that 
modifying only the argument of the isotropic distribution function in Eq.~(\ref{dist})
simply removes particles/modes with momentum components along the anisotropy direction,
i.e.~the plasma gets diluted with growing $\xi$. The momentum distribution defines the thermal parts of the bare propagators in the real-time formalism and 
thus parametrises a gas of non-interacting particles. Hence, the anisotropy must not affect the average number density of particles. This results in a normalisation condition
for all $\xi$ \cite{Romatschke:2003vc,Romatschke:2004jh},
\be
n=\int \frac{\mathrm{d}^3\bm{p}}{(2\pi)^3}f_\mathrm{iso}(\bm{p})\stackrel{!}{=}
\int \frac{\mathrm{d}^3\bm{p}}{(2\pi)^3}f(\bm{p},\xi)\;,
\ee
with the solution $N(\xi)=\sqrt{1+\xi}$.
For example, the integral over the Bose distribution function, 
$f_{\mathrm{iso}}(\bm{p})=n_B(\bm{p})$, 
evaluates to $n=3.606\,T^3/(2\pi)^2$. Using $N(\xi)=1$ for $\xi\neq 0$ gives
instead $n=\{2.55,1.09,0.3\}\,T^3/(2\pi)^2$ for $\xi=\{1,5,100\}$, respectively. 
Even for ``small'' anisotropy $\xi=1$, the dilution amounts to a significant 30\% and affects observables. 
We find that the modified 
normalisation largely compensates the effects observed previously,  
such that for $\xi=1$ there is little difference in the static potential 
between the isotropic and anisotropic cases. 
We then proceed to study the `maximal anisotropy effect' by considering the limit 
$\xi\rightarrow \infty$,
in which the plasma is confined to a thermal plane in momentum space 
perpendicular to the symmetry axis of the system. 
Somewhat surprisingly, also in this case the quarkonium spectral function is qualitatively in agreement with the isotropic situation. For simplicity a purely gluonic system is considered, the 
inclusion of light fermions is straightforward and will merely introduce a modification 
to self energy prefactors \cite{Arnold:2002zm}. 

\section{The static potential, propagator and self energy}

For our calculations we use the Keldysh-Schwinger real time formalism \cite{Lifshitz}, similar to \cite{Dumitru:2009fy}. There one considers matrices of correlation functions, $C_{ij}$, with indices
$i,j\in \{1,2\}$ denoting the forward and backward portions of the Keldysh contour. 
The real time static potential in a thermal medium may be defined through the quarkonium
Wightman function in the infinite quark mass limit, i.e.~the Wilson loop
in Minkowski time evaluated in a thermal bath \cite{Laine:2006ns} , whose time evolution
is governed by a `Schr\"odinger equation':
\be
\mathrm{i}C_{21}(t,\bm{r})=\frac{1}{N}\mathrm{Tr}<\hspace{-0.1cm}\begin{minipage}{2.25cm}
\centering\includegraphics[width=2cm]{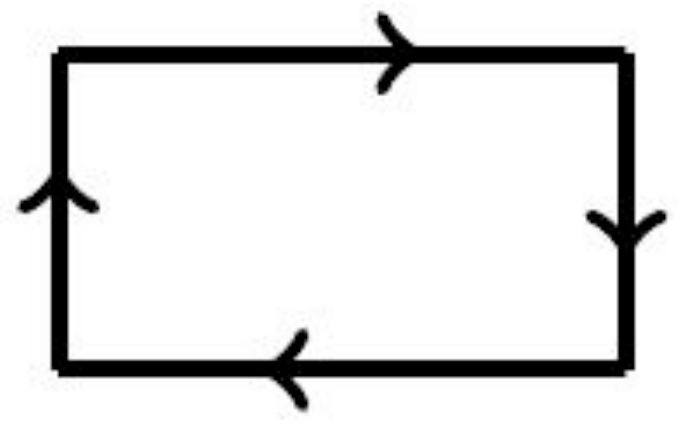}
\end{minipage}\hspace{-0.1cm}>\;,\qquad
\mathrm{i}\partial_t C_{21}(t,\textbf{r})=V(t,\bm{r})\,C_{21}(t,\textbf{r})\;.
\label{schr}
\ee
The static potential corresponds to the large time limit of this evolution, where the Wightman function
may be replaced by $C_{11}$,
\be\label{Vr}
V(\bm{r})=\lim_{t\rightarrow\infty} V(t,\bm{r})=\lim_{t\rightarrow\infty} 
\frac{\mathrm{i}\partial_t C_{11}(t,\textbf{r})}{C_{11}(t,\textbf{r})}\;,
\ee
and to leading order in perturbation theory is equivalent to a matching coefficient appearing
in a non-relativistic effective theory for finite temperatures  \cite{Escobedo:2008sy,Brambilla:2008cx,Beraudo:2007ky}.
Calculating the Feynman diagrams corresponding to the Wilson loop through order $g^2$
in the real time formalism, one finds the generic expression for the potential in 
an arbitrary (non-)equilibrium system \cite{Tassler:2008mg},
\ba
V(\bm{r})&=&g^2\mathrm{C_F}\int\frac{\mathrm{d}^3\bm{k}}{(2\pi)^3}
(1-\mathrm{e}^{\mathrm{i}\bm{k}\cdot\bm{r}})\tilde{G}_{11}^{00}(\omega=0,\bm{k})\nonumber \\
&=&
g^2\mathrm{C_F}\int\frac{\mathrm{d}^3\bm{k}}{(2\pi)^3}
(1-\mathrm{e}^{\mathrm{i}\bm{k}\cdot\bm{r}})\left(\frac{1}{2}\tilde{G}_\mathrm{S}^{00}+{\rm Re} 
\tilde{G}_\mathrm{R}^{00}\right)\;.
\label{RealTimeStaticPotential}
\ea
The tilde indicates a HTL-resummation of the gluon propagator.
Thus we need to evaluate the resummed retarded and symmetric propagators for the various degrees of anisotropy we wish to discuss. The propagators are solutions to the Dyson-Schwinger equations,
\ba
\tilde{\bm{G}}_\mathrm{R}&=& \bm{G}_\mathrm{R}+\bm{G}_\mathrm{R}\cdot\bm{\Pi}_{\mathrm{R}}\cdot\tilde{\bm{G}}_{\mathrm{R}},\nonumber \\
\tilde{\bm{G}}_{\mathrm{S}}&=&\bm{G}_\mathrm{S}+\bm{G}_{\mathrm{R}}\cdot\bm{\Pi}_{\mathrm{R}}\cdot\tilde{\bm{G}}_{\mathrm{S}}
+\bm{G}_{\mathrm{S}}\cdot\bm{\Pi}_{A}\cdot\tilde{\bm{G}}_{\mathrm{A}}
+\bm{G}_{\mathrm{R}}\cdot\bm{\Pi}_{\mathrm{S}}\cdot\tilde{\bm{G}}_{\mathrm{A}}\;.
\label{SchwingerDyson}
\ea
The bold face notation indicates matrices in Lorentz and colour space. In the following all colour 
matrices are $\sim \delta_{ab}$ and will be suppressed to simplify the notation.
The retarded self energy for an anisotropic plasma has been calculated in HTL-approximation in \cite{Mrowczynski:2000ed}, and the retarded propagator in a general covariant gauge in 
\cite{Dumitru:2007hy}. The resummed retarded gluon propagator in a general covariant gauge can be written in the form
\be
\tilde{G}_R^{\mu\nu}(K)=\Delta_A[A^{\mu\nu}-C^{\mu\nu}]+\Delta_G\left[(K^2-\alpha-\gamma)\frac{\omega^4}{K^4}B^{\mu\nu}+(\omega^2-\beta)C^{\mu\nu}+\delta\frac{\omega^2}{K^2}D^{\mu\nu}\right]-\frac{\lambda}{K^4}K^{\mu}K^{\nu}\;.
\label{prop}
\ee
Here the symmetry vector $n^\mu=(0,\bm{n})$ represents the anisotropy,
$K=(\omega,\bm{k})$,  $\lambda$ is the gauge parameter and the electric and magnetic propagators
are
\ba
\Delta_G^{-1}&=&(K^2-\alpha-\gamma)(\omega^2-\beta)-\delta^2[\bm{k}^2-(n\cdot K)^2]\;,\nonumber\\
\Delta_A^{-1}&=&K^2-\alpha\;.
\label{deltas}
\ea
Explicit expressions for 
the tensors $A-D$ and the structure functions
$\alpha(K,\xi)-\delta(K,\xi)$ for arbitrary anisotropies are listed in Eqs.~(2--5) in \cite{Dumitru:2007hy} 
and Eqs.~(B1--B4) in \cite{Romatschke:2003ms}, respectively.
Note however, that the structure functions given there are for $N(\xi)=1$ and need to be multiplied by $N(\xi)=\sqrt{1+\xi}$ for our purposes.  
Since we are mostly interested in applications for potential models, it is expedient
to take the static limit $\omega\rightarrow 0$ already at this stage 
to facilitate the calculations. In this case the bare symmetric propagator vanishes
and, using the fact that the retarded and advanced propagators are related via $\tilde{G}_{A}^{\mu\nu}=(\tilde{G}_{R}^{\nu\mu})^*$ in a general homogeneous system \cite{Lifshitz}, the second of Eqs.~(\ref{SchwingerDyson}) simplifies to
(see also \cite{Arnold:2002zm}) 
\be\label{DS}
\tilde{\bm{G}}_{\mathrm{S}}=\tilde{\bm{G}}_{\mathrm{R}}\cdot\bm{\Pi}_{\mathrm{S}}\cdot\tilde{\bm{G}}_{\mathrm{R}}^*\;.
\ee
The propagator in the static limit is (with the notation $k^2\equiv \bm{k}^2, 
\cos\theta_k\equiv \bm{n}\cdot\bm{k}/k$)
\ba
\tilde{G}_\mathrm{R}^{\mu\nu}(\omega=0,\bm{k})
&=&\Delta_\mathrm{A}[C^{\mu\nu}-A^{\mu\nu}]+\Delta_\mathrm{G}\left[(k^2+m^2_{\alpha}+m^2_{\gamma})B^{\mu\nu}-(k^2+m^2_{\beta})C^{\mu\nu}+\mathrm{i} \frac{m_{\delta}^2}{\sin{\theta_k}} D^{\mu\nu}\right]
\nonumber \\
&&-\frac{\lambda}{k^4}K^{\mu}K^{\nu}
\ea
with now
\be
\Delta_\mathrm{A}^{-1}=k^2+m^2_{\alpha},\hspace{1.0cm}\Delta_\mathrm{G}^{-1}=(k^2+m^2_{\alpha}+m^2_{\gamma})(k^2+m^2_{\beta})-m_{\delta}^4\;.
\ee
The tensor basis is simplified and takes the form
\be
\begin{tabular}{ccc}
$A^{0\mu}(\bm{k})=A^{\mu 0}(\bm{k})=0$&$A^{ij}(\bm{k})=\delta_{ij}-k^ik^j/k^2$&$B^{\mu\nu}(\bm{k})=\delta_{\mu 0}\delta_{\nu 0}$\\[0.1cm]
$C^{0\mu}(\bm{k})=C^{\mu 0}(\bm{k})=0$&$C^{ij}(\bm{k})=\tilde{n}^i\tilde{n}^j$&$D^{\mu\nu}(\bm{k})=\delta_{\mu 0}\tilde{n}^{\nu}+\delta_{\nu 0}\tilde{n}^{\mu}$
\end{tabular}
\ee
with $\tilde{x}^i=A^{ij}x^j$. A rescaling of the tensors 
$\frac{\omega^2}{k^2}B^{\mu\nu}\rightarrow B^{\mu\nu}$,
$\frac{\omega}{k^2}D^{\mu\nu}\rightarrow D^{\mu\nu}$, as well as 
$-\frac{\omega^2}{k^2}\Delta_G \rightarrow \Delta_G, -\Delta_A\rightarrow \Delta_A$ 
has been introduced with 
respect to Eqs.~(\ref{prop},\ref{deltas}) to keep all components of the retarded propagator 
explicitly finite in the static limit.
The effective masses are the static limits of the structure functions \cite{Romatschke:2003ms}, 
\be
m^2_{\alpha}=\lim_{\omega\rightarrow 0}\alpha\;, \quad
m^2_{\beta}=\lim_{\omega\rightarrow 0}-\frac{k^2}{\omega^2}\beta\;,\quad
m^2_{\gamma}=\lim_{\omega\rightarrow 0}\gamma\;,\quad
m^2_{\delta}=\lim_{\omega\rightarrow 0}\frac{k^2\sin{\theta_k}}{\omega}\mathrm{Im}(\delta)\;.
\ee

\section{Weak anisotropy, $\xi\leq 1$}

\begin{figure}[t]
\includegraphics[width=0.5\textwidth]{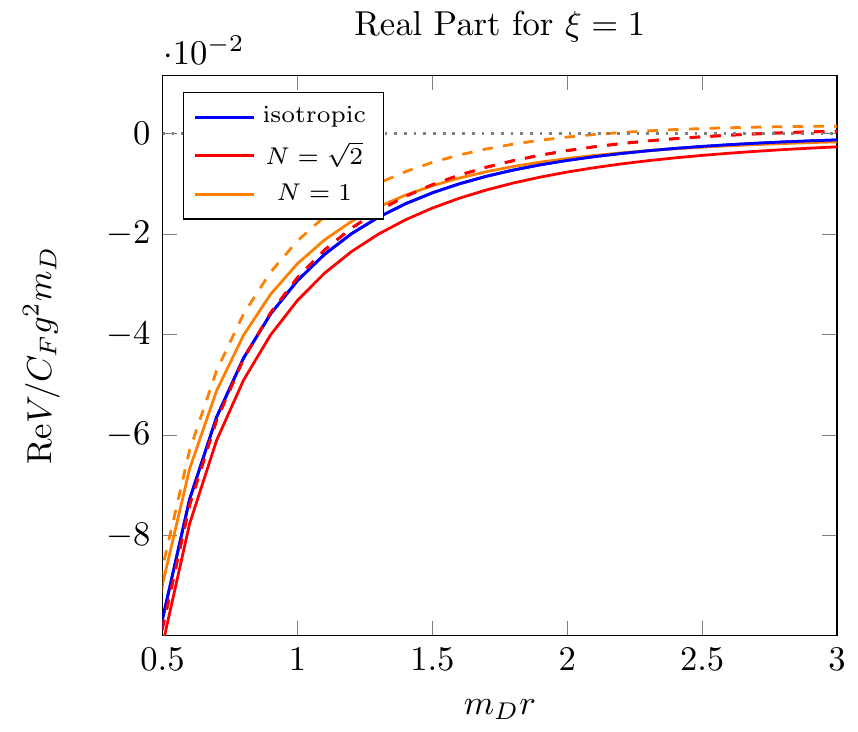}
\includegraphics[width=0.5\textwidth]{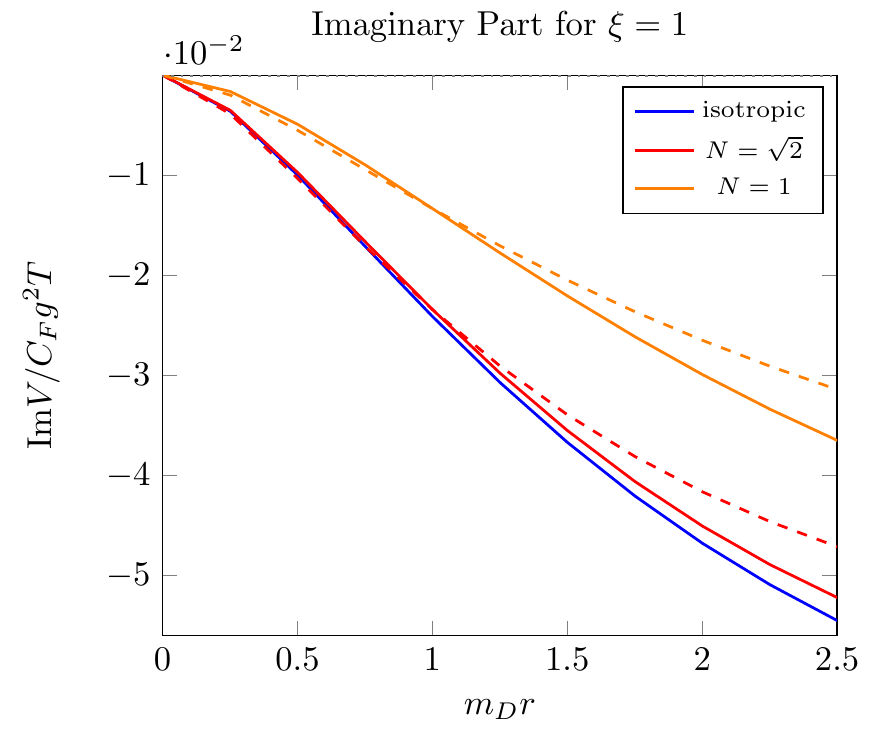}
\caption[]{Real (left) and imaginary (right) part of the static potential. The
blue curve corresponds
to the isotropic plasma, $\xi=0$, the other curves to $\xi=1$ with normalisation factor $N(\xi)=1$
(orange)
and $N(\xi)=\sqrt{1+\xi}$ (red). The potential is evaluated for $\theta_r=\frac{\pi}{2}$ (solid) and $\theta_r=0$ (dashed).
\label{smallxi}}
\end{figure}

In order to discuss the effect of a weak anisotropy, we employ the distribution function
Eq.~(\ref{dist}), including the non-trivial normalisation with $N(\xi)$, and expand the resummed propagators
up to linear order in $\xi$. For the longitudinal component of the retarded propagator appearing in Eq. (\ref{RealTimeStaticPotential}) we then have
\be
\tilde{G}_\mathrm{R}^{00}(\omega=0,\bm{k})=
\frac{k^2+m_\alpha^2+m_\gamma^2}{(k^2+m_\alpha^2+m_\gamma^2)(k^2+m_\beta^2)-m_\delta^4}\;.
\ee
All $\xi$-dependence resides in the effective masses. To linear order the 
dimensionless combinations $\hat{m}_x^2=m_x^2/m_D^2$ read 
\ba
\hat{m}_\alpha^2=-\frac{\xi}{3}\cos^2\theta_k,
&&\hat{m}_\beta^2=1+\xi(\cos^2 \theta_k-\frac{1}{6}),\\
\hat{m}_\gamma^2=\frac{\xi}{3}\sin^2\theta_k,
&&\hat{m}_\delta^2=-\xi\frac{\pi}{4}\sin{\theta_k}\cos{\theta_k}\;.
\ea
Note that $\hat{m}_\beta^2$ differs from the expression given in 
\cite{Romatschke:2003ms} due to the normalisation factor $N(\xi)=\sqrt{1+\xi}$
in the distribution function Eq.~(\ref{dist}).
The retarded propagator to linear order then is
\be
\tilde{G}_\mathrm{R}^{00}=\frac{1}{k^2+m_D^2}-\xi\frac{m_D^2}{(k^2+m_D^2)^2}
(\cos^2\theta_k-\frac{1}{6})\;.
\ee
Expanding also the static potential, $V(\bm{r})=V_0(r)+\xi V_1(\bm{r})+\ldots$, 
we find for the correction of the real part due to the anisotropy,
\be
{\rm Re} V_1(\bm{r})=g^2C_F\,m_D^2\int\frac{\mathrm{d}^3\bm{k}}{(2\pi)^3}\left(1-\cos(\bm{k}\cdot\bm{r})\right)\frac{1}{(k^2+m_D^2)^2}\left(\cos^2\theta_k-\frac{1}{6}\right).
\ee
For the imaginary part we need in addition the symmetric propagator and the symmetric
self-energy, cf.~Eq.~(\ref{DS}). In the static limit and for soft external momenta $k_i\ll p_i$ the latter takes the form \cite{Arnold:2002zm}
\be
\mathrm{i}\Pi_\mathrm{S}^{\mu\nu}=8\pi g^2 N\frac{1}{k}\int \frac{d^3p}{(2\pi)^3}v_p^{\mu}v_p^{\nu}f(\bm{p})(1+f(\bm{p}+\bm{k}))\delta(\bm{v}_p\cdot\bm{v}_k)\;.
\label{self}
\ee
Straightforward calculation gives the following expansion for the symmetric
propagator 
\ba \tilde{G}_\mathrm{S}^{00}&=&-\mathrm{i}\frac{2\pi}{k(k^2+m_D^2)^2}\frac{m_D^2}{\beta}
\left(1+\xi\left[\frac{\pi^2-3\zeta(3)}{\pi^2}-\frac{3}{4}\sin^2\theta_k\right]\right)
\nonumber\\
&&+i\xi 4\pi\frac{m_D^4}{\beta}\frac{1}{k(k^2+m_D^2)^3}(\cos^2\theta_k-\frac{1}{6})\;.
\ea
where $\zeta$ is the Riemann zeta function. For $N(\xi)=1$ the first fraction in square brackets
is absent  and $1/6\rightarrow 2/3$ in the last line. 
In this case we reproduce the symmetric propagator from \cite{Dumitru:2009fy}. 
The leading correction to the imaginary part of the
potential due to the anisotropy is given by
\be
{\rm Im} V_1(r)=g^2C_F\int \frac{\mathrm{d}^3\bm{k}}{(2\pi)^3}(1-\cos \bm{k}\cdot\bm{r})
\frac{1}{2}\frac{\partial}{\partial \xi}\tilde{G}_S^{00}\Big|_{\xi=0}
\ee
The remaining integrations can be done numerically, and the results for the real and imaginary parts of the static potential are shown in Fig.~\ref{smallxi}, where we have defined the angle between
$\bm{r}$ and the anisotropy axis, $\cos\theta_r\equiv \bm{r}\cdot\bm{n}/r$.
For the normalisation $N(\xi)=1$ it has been observed that an increasing anisotropy
lowers the real part of the static potential towards a Coulomb potential \cite{Dumitru:2007hy}. 
However, this follows trivially for $\xi \rightarrow \infty$ since the medium is 
disappearing in this case, and hence $m_D\rightarrow 0$. By contrast, the effect of the anisotropy 
is much weaker when the normalisation $N(\xi)=\sqrt{1+\xi}$ is chosen. Note also, that in this
case the potential is even slightly enhanced at small distances.
The difference between the two normalisations is even more pronounced in the imaginary
part, which is caused by Landau damping and introduces a broadening of the spectral function \cite{Laine:2007gj}.
A weaker damping effect is seen for $N(\xi)=1$, leading to the conclusion that
bound states should melt at higher temperatures \cite{Dumitru:2009ni,Burnier:2009yu,Dumitru:2009fy}. 
For $N(\xi)=\sqrt{\xi+1}$, this effect is modified and persists only
at very large distances. For smaller distances, on the scale of the bound states, there is only very little difference to the isotropic case. Uncertainties due to non-perturbative corrections to the diffusive physics underlying the existence of the imaginary part were found to be significantly larger in recent studies \cite{Laine:2007qy,Laine:2009dd}. 

\section{Strong anisotropy, $\xi\rightarrow \infty$}

Having seen that, with a distribution function normalised to maintain particle number in an ideal gas, 
the effects of a weak momentum space anisotropy on the static potential are rather small, we are now asking how a maximally anisotropic plasma would behave. We begin by returning to the gluon propagator 
in a general covariant gauge, Eq.~(\ref{prop}), and discuss some aspects of the dispersion relations 
which have been derived in \cite{Romatschke:2004jh}.

\subsection{Dispersion relations}

In the limit of maximal anisotropy, the structure functions $\alpha(K)-\delta(K)$ take on an analytic form \cite{Romatschke:2004jh}, facilitating a numerical evaluation of the dispersion relations of the retarded gluon propagator,
$\Delta_G=0$ and $\Delta_A=0$. The associated modes are referred to as electric and magnetic modes respectively (for an analysis of dispersion relations at arbitrary anisotropies see 
\cite{Romatschke:2003ms,Romatschke:2004jh}). They can be conveniently expressed by introducing effective pole masses $m(\bm{k})$, which satisfy the relation
\be
\omega^2=k^2+m^2(\bm{k})
\ee
at poles and branch cuts of the electric and magnetic propagators.
\begin{figure}[t]
\begin{center}
\hspace{-5mm}\includegraphics[width=7.5cm]{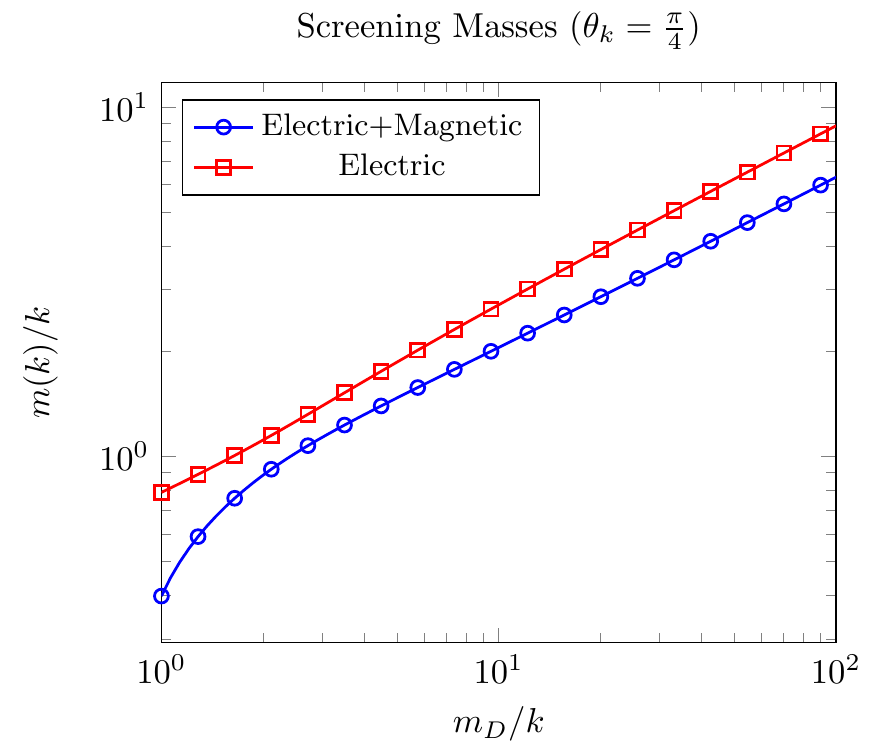}\includegraphics[width=7.5cm]{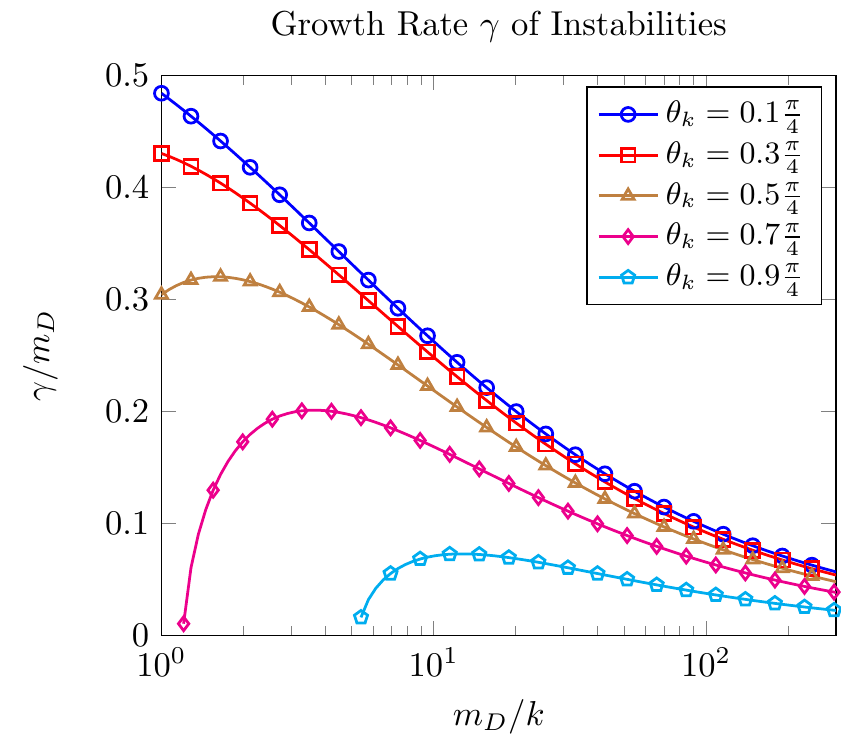}
\caption[]{The effective masses of stable electric and magnetic modes (shown on the left for $\theta_k=\pi/4$ and $k\leq m_D$) indicate a screening of long range interactions. For $\theta_k<\pi/4$ the gluon propagator has poles on the imaginary $\omega$-axis due to the Weibel instability. The growth rate $\gamma=\mathrm{Im}(\omega)$ of the associated unstable modes (right) increases towards the symmetry axis.
\label{dispersionplot}}
\end{center}
\end{figure}
The results from a numerical determination of the pole masses are illustrated in Fig.~\ref{dispersionplot} and summarised in the following. The dynamical screening masses of the lowest electric mode as well as the magnetic mode are found to be comparable in magnitude to the screening masses in an isotropic medium with a moderate angular dependence of both masses towards higher momenta. The findings indicate that strongly coupled interactions over large distances are suppressed leading to a deconfinement of colour charges. This is a prerequisite for the possibility of a consistent perturbative treatment using the HTL approach.

For $\theta_k<\pi/4$ the gluon propagator has additional poles along the imaginary $\omega$-axis indicating the existence of unstable field modes. The instability, which is present in a general anisotropic QCD plasma, is speculated to be responsible for the rapid isotropisation of the quark gluon plasma observed in heavy ion collisions \cite{Mrowczynski:1993qm}. The growth rate $\gamma=\mathrm{Im}(\omega)$ of unstable modes is observed to increase and to peak at higher momenta towards the symmetry axis.

\subsection{The static potential}

Next, we take the static limit of the maximally anisotropic propagator 
in order to discuss the static potential. At maximal anisotropy
the appropriate effective masses read:
\be\label{masses}
m^2_{\alpha}=-\frac{\pi m_D^2}{4 \tan^2 \theta_k}\;,\quad
m^2_{\beta}=\frac{\pi m_D^2}{4}\;,\quad
m^2_{\gamma}=\frac{\pi m_D^2}{4} \left(1+\frac{2}{\tan^2 \theta_k}\right)\;,\quad
m^2_{\delta}=\frac{\pi m_D^2}{4}\frac{\cos \theta_k}{\sin^2 \theta_k}\;.
\ee
We also need to re-evaluate the symmetric self energy, Eq.~(\ref{self}). 
Since the momentum distribution function for the anisotropic medium $f(\bm{p})$ becomes sharply peaked around $\bm{v}_p\cdot\bm{n}=0$ upon sending $\xi$ to infinity, the following identity may be derived \cite{Arnold:2003rq}:
\be
\lim_{\xi\rightarrow\infty} f(\bm{p}) = \delta(\bm{v}_p\cdot\bm{n})\int_{-\infty}^{\infty}dx \; n_\mathrm{B}(p\sqrt{1+x^2})=\delta(\bm{v_p}\cdot\bm{n})h(p).
\ee
The self energy thus takes the following analytically tractable form:
\be
\mathrm{i}\Pi_{\mathrm{S}}^{\mu\nu}(\bm{k})=\frac{8\pi g^2N}{k}\int \frac{\mathrm{d}^3\bm{p}}{(2\pi)^3} \hspace{0.1cm}v_p^{\mu} v_p^{\nu}\; h(p)\delta(\bm{v}_p\cdot\bm{n})\delta(\bm{v}_p\cdot\bm{v}_k)
\left(1+h(p)\frac{|\bm{p}+\bm{k}|}{k}\delta(\bm{v}_k\cdot\bm{n})\right).
\ee
The Bose enhanced part of the self energy is amplified by a factor $|\bm{p}+\bm{k}|/k\approx p/k$ which will translate into an amplification by $T/k$, indicating that scattering processes are dominated by Bose enhanced scattering into the thermal plane. The self energy is diagonal, 
\be
\mathrm{i}\Pi_{\mathrm{S}}^{00}=\mathrm{i}\Pi_{\mathrm{S}}^{jj}=\frac{m_D^2}{k\beta\sin\theta_k}\left(c_{\parallel}+\frac{c_{\perp}}{k\beta}\delta(\bm{v}_k\cdot\bm{n})\right),~~~k_i \Pi_{\mathrm{S}}^{ij} = n_i \Pi_{\mathrm{S}}^{ij} =0\;,
\ee
with $m_D^2=\frac{N}{3} g^2T^2$ and the prefactors $c_x$ are obtained from the following integrals
\ba
c_{\parallel}&=&\frac{6}{\pi^2}\beta^3\int_0^{\infty}dp \;p^2 h(p)=2.9231\ldots\nonumber\\
c_{\perp}&=&\frac{6}{\pi^2}\beta^4\int_0^{\infty}dp \;p^3 h^2(p)=3.4230\ldots
\ea
The symmetric propagator can now be calculated from (\ref{DS}) in a straightforward fashion. Its longitudinal component can be separated in contributions dominated 
by external momenta perpendicular and parallel to the symmetry axis
\ba\label{GS}
\mathrm{i}\tilde{G}_S^{00}(\bm{k})&=&\mathrm{i}\tilde{G}_S^{\perp}(\bm{k})+\mathrm{i}\tilde{G}_S^{\parallel}(\bm{k}) \nn
\mathrm{i}\tilde{G}_S^{\perp}(\bm{k})&=&\frac{c_{\perp} m_D^2}{\beta^2} \,\frac{1}{k^2(k^2+\frac{\Pi}{4}m_D^2)^2}\delta(\cos\theta_k)\nonumber \\
\mathrm{i}\tilde{G}_S^{\parallel}(\bm{k})&=&\frac{c_{\parallel} m_D^2}{\beta k\sin{\theta_k}} \,\frac{(k^2+m_{\alpha}^2+m_{\gamma}^2)^2}{[(k^2+m_{\alpha}^2+m_{\gamma}^2)(k^2+m_{\beta}^2)-m^4_{\delta}]^2}\;.
\ea
The first component corresponds to Bose enhanced scattering into the thermal plane and is enhanced in $\frac{T}{k} \gg 1$ while the second component, corresponding to scattering away from the thermal plane, is suppressed even for $\sin\theta_k\rightarrow 0$ due to the diverging effective mass $m^2_{\delta}$ in this limit. Contributions from the symmetric propagator will therefore be dominated by momenta perpendicular to the symmetry axis,
\be
\mathrm{i}\tilde{G}_S^{00}(\bm{k})\approx \mathrm{i}\tilde{G}_S^{\perp}(\bm{k})\;.
\ee

\begin{figure}[t]
\begin{center}
\hspace{-5mm}\includegraphics[width=7.5cm]{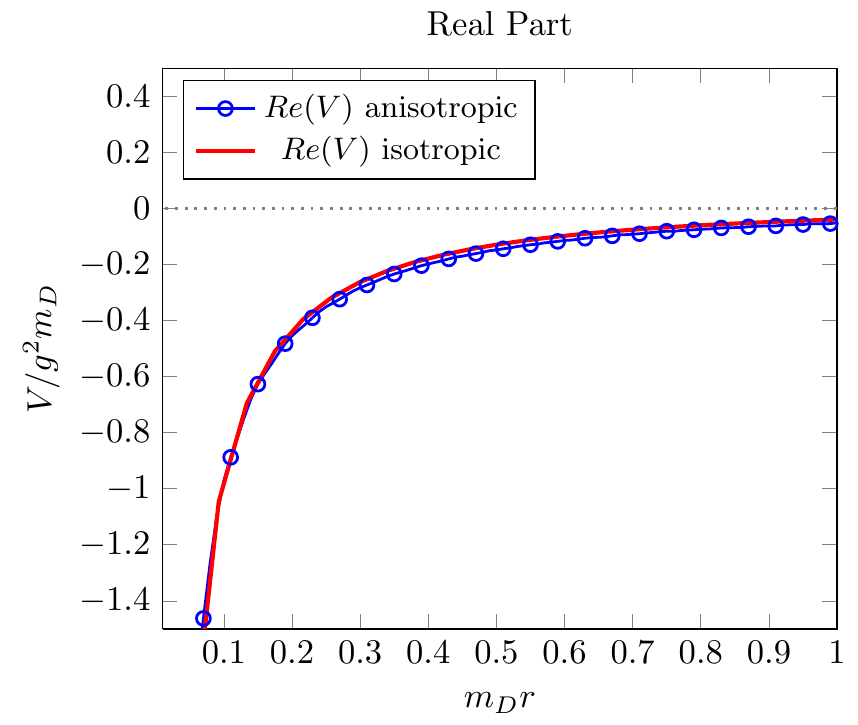}\includegraphics[width=7.5cm]{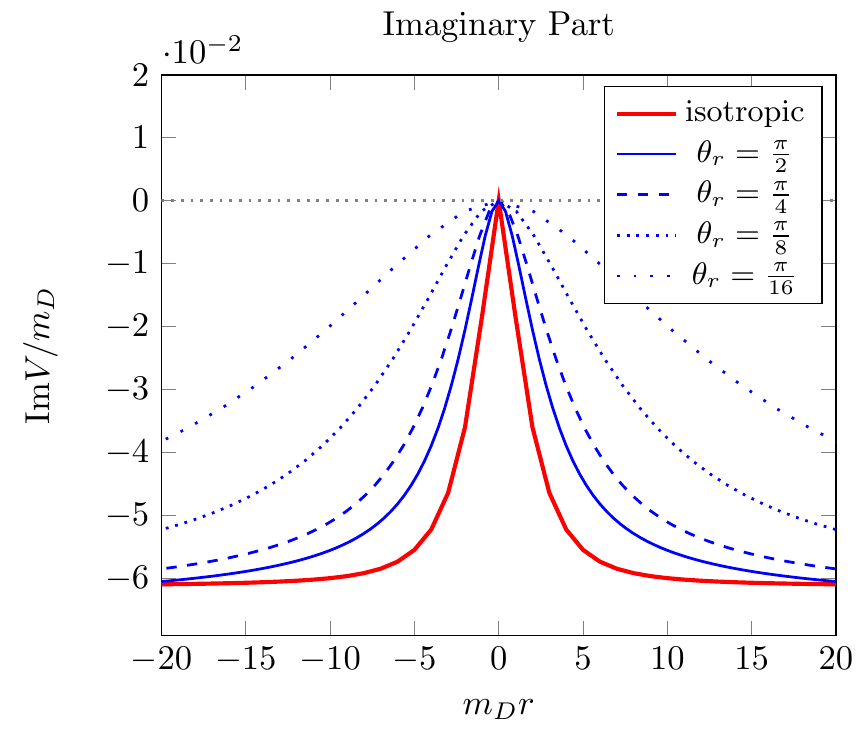}
\caption[]{The real part of the anisotropic static potential, normalized to $V(\infty)=0$ and the imaginary part shown as a function of the radial separation $r$ for $N=3$ and $g^2=3$. The cylindrical symmetry of the anisotropic imaginary part is illustrated by varying $\theta_r$.
\label{revplot}}
\end{center}
\end{figure}
\noindent
We are now ready to discuss the static potential in a maximally anisotropic medium, by inserting
these ingredients into Eq.~(\ref{RealTimeStaticPotential}).
Upon choosing a spherical coordinate system with the anisotropy vector $\bm{n}$ pointing in the z-direction, $\cos\theta_r\equiv \bm{r}\cdot\bm{n}/r$ and with the abbreviation $x=\cos\theta_k$, 
one angular integration can be performed immediately leading to 
\ba
\mathrm{Re}V(\bm{r})&=&g^2\mathrm{C_F}\int \frac{\mathrm{d}k \,\mathrm{d}x}{(2\pi)^2}\left(1-\mathrm{J_0}
(kr\sqrt{1-x^2}\sin{\theta_r} )\cos{(krx\cos{\theta_r})}\right)\nonumber\\
&&\hspace{2.5cm}\times \frac{k^2(k^2+m^2_{\alpha}+m^2_{\gamma})}{(k^2+m^2_{\alpha}+m^2_{\gamma})(k^2+m^2_{\beta})-m_{\delta}^4}\;,
\ea
where $\mathrm{J_n}$ is the Bessel function of the first kind. The expression is readily evaluated numerically using standard integration routines \cite{Hahn:2004fe}. The potential contains a linear 
divergence which we subtract by normalising $V(|\bm{r}|\rightarrow\infty)=0$. The resulting real part displays an approximate spherical symmetry. The radial profile, as depicted in 
Fig.~\ref{revplot} for $N=3$, turns out to be in good agreement with the standard Debye screened potential. For the
imaginary part we obtain in a similar way
\be
\mathrm{Im}V(\bm{r})=-\frac{3 \mathrm{C_F} m_D^4 c^{\perp}}{8 N \pi^2}\int \mathrm{d}k \;(1-\mathrm{J_0}(k r \sin{\theta_r}))\frac{1}{(k^2+\frac{\pi}{4}m_D^2)^2}\;.
\ee
The imaginary part thus only depends on the separation $r \sin{\theta_r}$ of the static quark pair in the thermal plane perpendicular to the anisotropy vector. Unlike the previously studied quantities the imaginary part turns out to be quantitatively different from its isotropic counterpart \cite{Laine:2006ns}. 
The damping of the static quark correlator is significantly weakened towards the anisotropy axis for
reasonable values of the coupling.

\subsection{Quarkonium spectral function}

\begin{figure}[t]
\begin{center}
\hspace{-0.5cm}\includegraphics[width=7.5cm]{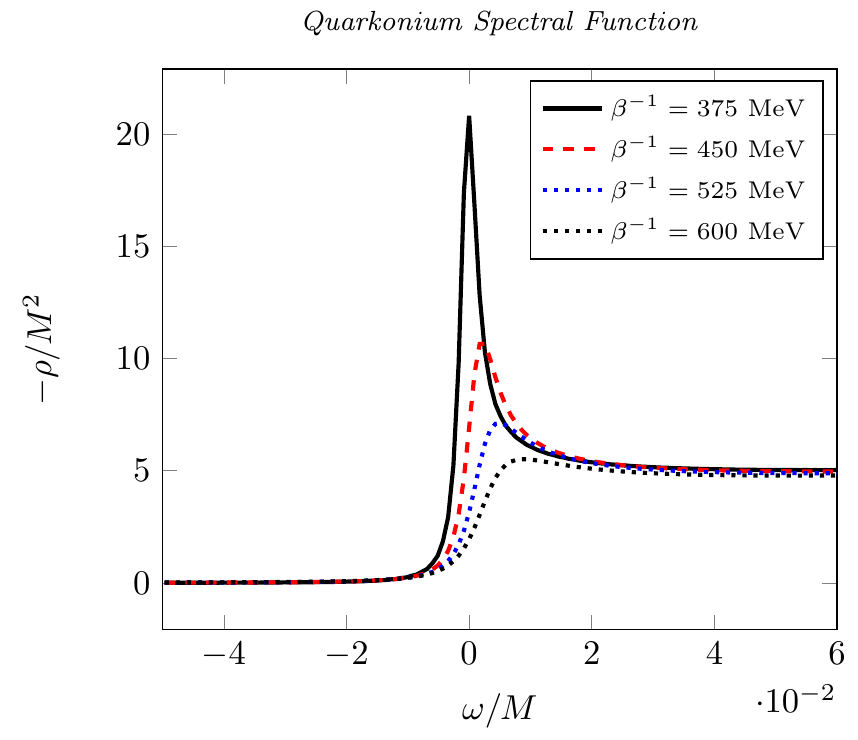}\includegraphics[width=7.5cm]{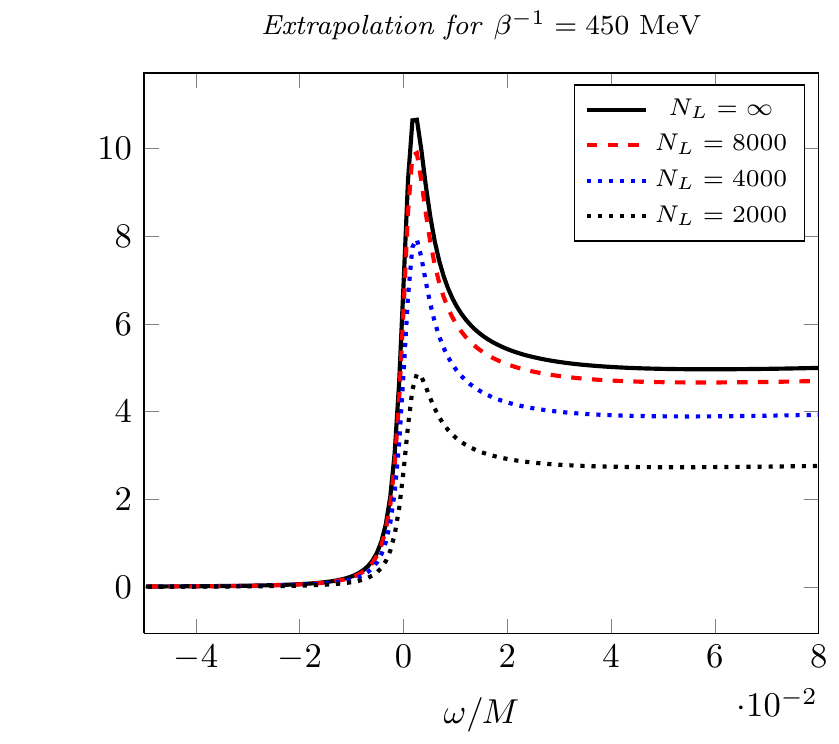}\\
\caption[]{Left: The melting of the bottomonium spectral function in a maximally anisotropic quark gluon plasma at a fixed coupling of $g^2=3$ ($\alpha_s\approx 0.25$). Right: The polynomial extrapolation of the spectral function using three different grid spacings.
\label{meltingplot}}
\end{center}
\end{figure}

In order to compare a more physical quantity between the isotropic and anisotropic cases,
we also calculate the quarkonium spectral function.
Since this has not been done before for anisotropies, we present it in some detail. For a
$q\bar{q}$ system with finite constituent mass $M$ the following Schr\"odinger
equation must be solved \cite{Laine:2006ns},
\be\label{Schroedinger}
\mathrm{i}\partial_t
C_{21}(t,\bm{r})=\left(2M-\frac{\triangle_{\bm{r}}}{M}+V(\bm{r})\right)C_{21}(t,\bm{r}),\hspace{1cm}\mathrm{i}C_{21}(0,\bm{r})=-6N\delta^{(3)}(\bm{r})\;
\ee
where $V(\bm{r})$ is the real time static potential discussed in the previous
sections. For $\bm{r}=\bm{0}$ the quarkonium correlator coincides
with the quark current correlator $C_{21}(t,\bm{0})=\langle
J^{\mu}(t)J_{\mu}(0)\rangle$, where $J_{\mu}=\overline{\psi}\gamma_{\mu}\psi$.
The quark pair is assumed to be created at time $t=0$ and the initial
condition for the Schr\"odinger equation (\ref{Schroedinger}) ensues from the
afore mentioned relation.
The spectral function for quarkonium at rest $\rho(\omega)$ can  be calculated from the difference of the 
two off-diagonal elements of the heavy-quarkonium correlator,
\be
\rho(\omega)=\int \mathrm{d}t \;\mathrm{e}^{-\mathrm{i}\omega t}\left\{\mathrm{i}C_{21}(t,\mathbf{0})-\mathrm{i}C_{12}(t,\mathbf{0})\right\}\;.
\ee
In thermal equilibrium both components are related via the Kubo- Martin- Schwinger condition, 
which does not hold for the system under consideration. In the operator formalism the difference between the two correlators takes the form
\be
C_{21}(t,\mathbf{0})-C_{12}(t,\mathbf{0})=\frac{1}{Z}\mathrm{Tr} \left\{\hat{\sigma} \hat{J}_{\mu}(t,\mathbf{0})\hat{J}^{\mu}(0,\mathbf{0})\right\}-\frac{1}{Z}\mathrm{Tr} \left\{\hat{\sigma}\hat{J}_{\mu}(0,\mathbf{0})\hat{J}^{\mu}(t,\mathbf{0})\right\}\;,
\ee
where the trace is over all energy eigenstates and $\hat{\sigma}$ is the statistical operator of the anisotropic system. Using the cyclicity of the trace, $C_{12}$ is expressed as
\be
C_{12}(t,\mathbf{0})=\frac{1}{Z}\mathrm{Tr} \left\{\hat{J}_{\mu}(t,\mathbf{0})\hat{\sigma}\hat{J}^{\mu}(0,\mathbf{0})\right\}\;.
\ee
The difference between both contributions is that in $C_{12}$ the density matrix acts on the Hilbert space of the plasma with the quark pair present. For $2\beta M\gg 1$, where the quarkonium threshold $\simeq 2 M$ is much higher than the average particle momentum $\beta^{-1}$ in the thermal plane, the contribution of the correlator $C_{12}$ will be substantially suppressed. This justifies the following approximation which is also employed in the equilibrium case \cite{Laine:2007gj},
\be
\rho(\omega)\approx\int dt \;\mathrm{e}^{-\mathrm{i}\omega t}C_{21}(t,\mathbf{0}).
\ee
The Wightman propagator $C_{21}(t,\bm{r})$ for a static quark pair is obtained at large positive times as a solution of the Schr\"odinger Eq.~(\ref{Vr}). 
Similarly, using $\mathrm{i}C_{ij}(-t,\bm{r})=[\mathrm{i}C_{ij}]^*(t,\bm{r})$, the static potential in the infinite past is
\be
V'(\bm{r})=\lim_{t\rightarrow-\infty}\frac{\mathrm{i}\partial_t C_{22}(t,\bm{r})}{C_{22}(t,\bm{r})}=-g^2\mathrm{C_F}\int\frac{\mathrm{d}^3\bm{k}}{(2\pi)^3}(1-\cos{\bm{k}\cdot\bm{r}})\tilde{G}_{22}^{00}=V^*(\bm{r})\;.
\ee
The spectral function can thus be computed from the relation
\be
\rho(\omega)=2\int_0^{\infty}dt\left\{\cos{(\omega t)}\mathrm{Re}(\mathrm{i}C_{21})-\sin{(\omega t)}\mathrm{Im}(\mathrm{i}C_{21})\right\}.
\ee
The numerical solution of the Schr\"odinger equation for the Wightman correlator $C_{21}$ is non-trivial. The task consists in the interesting problem to solve a parabolic complex differential equation in three dimensions with a potential consisting of a parametrically large spherically symmetric real part and a small cylindrically symmetric imaginary part. Due to the positive parity and the symmetry of the potential with respect to the anisotropy axis, the correlator may be expanded in a subset of spherical harmonics,
\ba
C_{21}(t,\bm{r})&=&c_l(t,r)\mathrm{Y}_{2l}^0(\theta_r)\;,\nonumber\\
V(\bm{r})&=&V^{(re)}(r)+\mathrm{i} V^{(im)}_l(r)\mathrm{Y}_{2l}^0(\theta_r)\;,
\ea
with $V^{(re)}(r)$ and $V_l^{(im)}(r)$ denoting the real and imaginary parts of the potential respectively. With the rescaling
\be
c_l(t,r)=\frac{u_l(t,r)}{r}\mathrm{e}^{-\mathrm{i}2Mt},
\ee
the one-dimensional Schr\"odinger equations for the components of the spherical expansion take a numerically tractable form,
\be\label{SchroedingerSystem}
\mathrm{i}\partial_t u_l=-\frac{1}{M}\left(\frac{\mathrm{d}^2}{\mathrm{d}r^2}-\frac{2l(2l+1)}{r^2}\right)u_l+V^{(re)}u_l+\mathrm{i}\mathcal{P}_{lmn}V_m^{(im)}u_n\;.
\ee
The tensor $\mathcal{P}_{lmn}$ is defined by a unit sphere integration over a product of three spherical harmonics which can be evaluated exactly via Wigner-3j symbols \cite{Arfken}: 

\be
\mathcal{P}_{lmn}=\int \mathrm{d}\Omega \mathrm{Y}_{2l}^0\mathrm{Y}_{2m}^0\mathrm{Y}_{2n}^0
=\sqrt{\frac{(4 l+1)(4 m +1)(4 n +1)}{4\pi}}\left(
\begin{array}{ccc}
2l&2m&2n\\
0&0&0
\end{array}
\right)^2\;.
\ee
Without the coupling between different spherical components, a relation of the following type needs to be evaluated for every timestep $\mathrm{d}t$,
\be
\tilde{u}_l(t+\mathrm{d}t)=\mathrm{e}^{-\mathrm{i}\mathrm{\hat{H}}_l \mathrm{d}t}u_l(t),
\ee
where the differential operator $\mathrm{\hat{H}}_l$ is given by
\be
\mathrm{\hat{H}}_l=-\frac{1}{M}\left(\frac{d^2}{dr^2}-\frac{2l(2l+1)}{r^2}\right)+V^{(re)}(r)+\mathrm{i}\mathcal{P}_{lml}V_m^{(im)}(r)\;.
\ee
To ensure a stable time evolution we employ the Crank-Nicolson scheme \cite{CrankNicolson} by introducing the second order approximation
\be
\left(1+\frac{1}{2}\mathrm{i}\mathrm{\hat{H}}\mathrm{d}t\right)\tilde{u}_l(t+\mathrm{d}t)=
\left(1-\frac{1}{2}\mathrm{i}\mathrm{\hat{H}}\mathrm{d}t\right)u_l(t)\;,
\ee
which is solved as a tridiagonal matrix equation for the discretized system. The full problem is solved without sacrificing the stability of the algorithm by treating the parametrically small coupling between the various spherical components as a linear perturbation,
\be
u_l(t+\mathrm{d}t)=\tilde{u}_l(t+\mathrm{d}t)+\mathrm{d}t \sum_{m,n\neq l}\mathcal{P}_{lmn}V^{(im)}_m\ u_n(t)\;.
\ee
The system is discretised on a finite lattice with spacing $a$ and lattice sites $[1,\ldots,N_L]$, choosing the boundary conditions
\be
u_l(0)=0~~~~\textrm{and}~~~~u_l(N_L+1)=0
\ee
for all spherical components with $l$ limited to the range $l \in [0,\ldots,l_{max}]$.
The initial condition (\ref{Schroedinger}) is discretised as follows \cite{Laine:2007gj},
\be
u_0(0,r)=-6N\frac{r}{4\pi}\delta(r)\rightarrow -6N\left(\frac{2}{\pi a}\right)^2\frac{n(-1)^{n+1}}{4 n^2 -1}\;,\nonumber
\ee
with $r=n a$ being the discretised radial coordinate. All other spherical components $u_{l\neq 0}(0,r)$ are set to zero.
The scale for the numerical determination of the bottomonium spectral function for $N=3$ shown in  
Fig.~\ref{meltingplot} is set by the QCD scale $\Lambda_{\overline{MS}}\simeq 300 MeV$ in the $\overline{MS}$ scheme. The bottom quark mass $M=14.15\Lambda_{\overline{MS}}$ is chosen in agreement with \cite{Laine:2007gj}. Introducing a spherical cutoff $l_{max}=10$ the system of coupled one dimensional Schr\"odinger equations is solved on three different grids $N_L\in\{2000,4000,8000\}$ with a fixed physical size of $a N_L=400/\Lambda_{\overline{MS}}$. A time-step of $\mathrm{d}t=0.02 a$ is chosen and in all three cases the system is evolved for a fixed time $t=265/\Lambda_{\overline{MS}}$. The spectral function depends linearly on the lattice spacing $a$ since the coupling between the different spherical components of (\ref{SchroedingerSystem}) is introduced as a linear perturbation. To take into account $a^2$-corrections, the spectral function is extrapolated to the continuum by fitting the $a$-dependence of the spectral function at a given frequency against a quadratic polynomial. 
The melting of the quarkonium resonance due to Landau damping effects is in qualitative agreement with the decay of the resonance in thermal equilibrium obtained using the same approach \cite{Laine:2007gj}. 

\section{Conclusions}
We have reconsidered the effects of an anisotropic momentum distribution on the real-time static potential of a heavy quark pair in a plasma using
a normalised distribution function which keeps particle numbers constant in the ideal gas limit.  
At small anisotropies, previously observed effects 
were shown to be mostly due to a particle dilution stipulated by a different normalisation factor. 
In order to probe for a larger effect we then looked at maximal anisotropy.
The calculation of the quarkonium spectral function indicates a melting of the quarkonium resonance similar to the one for an isotropic medium due to a screening of the Coulomb potential and a dissociation of bound states by Landau damping effects in the thermal plane. Strongly coupled interactions over large distances are suppressed due to the generation of nearly isotropic screening masses comparable in magnitude to their equilibrium counterparts. 
For small angles with the anisotropy axis, the gluon propagator has additional poles along the imaginary $\omega$-axis indicating the existence of unstable field modes. 
The growth rate $\gamma=Im(\omega)$ of unstable modes is observed to increase and to peak at higher momenta towards the symmetry axis.

The overall physical picture that emerges from a weak-coupling analysis of this system indicates that the strongly anisotropic state of matter created in the initial stages of heavy-ion collisions is characterised by an early deconfinement of colour charges and a rapid diffusive transport of these charges in the collision plane. The most striking result is that observables related to quarkonium are very similar to their counterparts in an isotropic medium, making it difficult to distinguish a highly anisotropic plasma from its isotropic counterpart on the basis of heavy quark observables.

\section*{Acknowledgements:}
We thank A.~Dumitru and P.~Romatschke for discussions and comments on the manuscript.
This work is supported by the German BMBF, project
{\em Hot Nuclear Matter from Heavy Ion Collisions
and its Understanding from QCD}, No.~06MS254.



\end{document}